\documentclass[12pt,preprint]{emulateapj}
\usepackage{apjfonts}
\usepackage[dvipdfm,colorlinks,linkcolor=red,anchorcolor=black,citecolor=blue]{hyperref}
\begin{document}

\title{The spatially-resolved NUV$-r$ color of local star-forming galaxies and clues for quenching}
\shortauthors{Pan et al.}
\shorttitle{Spatially-resolved NUV$-r$ color of star-forming galaxies}
\author{Zhizheng Pan\altaffilmark{1}, Xianzhong Zheng \altaffilmark{1}, Weipeng Lin \altaffilmark{2,3}, Jinrong Li\altaffilmark{4,5}, Jing Wang\altaffilmark{6},Lulu Fan\altaffilmark{7}, Xu Kong\altaffilmark{4,5} }

\email{panzz@pmo.ac.cn, xzzheng@pmo.ac.cn, linwp@shao.ac.cn, xkong@ustc.edu.cn}
\altaffiltext{1}{Purple Mountain Observatory, Chinese Academy of Sciences, 2 West-Beijing Road, Nanjing 210008, China}
\altaffiltext{2}{School of Astronomy and Space Science, Sun Yat-Sen University, Guangzhou, 510275, China}
\altaffiltext{3}{Key laboratory for research in galaxies and cosmology, Shanghai Astronomical Observatory,
Chinese Academy of Science, 80 Nandan Road, Shanghai, 200030, China}
\altaffiltext{4}{Center of Astrophysics, University of
Science and Technology of China, Jinzhai Road 96, Hefei 230026, China}
\altaffiltext{5}{Key Laboratory for Research in Galaxies and Cosmology, USTC,
CAS, China}
\altaffiltext{6}{CSIRO Astronomy \& Space Science, Australia Telescope National Facility, PO Box 76, Epping, NSW 1710, Australia}
\altaffiltext{7}{Institute of Space Sciences, Shandong University, Weihai, 264209, China}

\begin{abstract}
Using a sample of $\sim$6,000 local face-on star-forming galaxies (SFGs), we examine the correlations between the NUV$-r$ colors both inside and outside the half-light radius, stellar mass $M_{\ast}$ and S\'{e}rsic index $n$ in order to understand how the quenching of star formation is linked to galaxy structure. For these less dust-attenuated galaxies, NUV$-r$ is found to be linearly correlated with $D_{n}4000$, supporting that NUV$-r$ is a good photometric indicator of stellar age (or specific star formation rate). We find that: (1) At $M_{\ast}<10^{10.2}M_{\sun}$, the central NUV$-r$ is on average only $\sim$ 0.25 mag redder than the outer NUV$-r$. The intrinsic value would be even smaller after accounting for dust correction. However, the central NUV$-r$ becomes systematically much redder than the outer NUV$-r$ for more massive galaxies at $M_{\ast}>10^{10.2}M_{\sun}$. (2) The central NUV$-r$ shows no dependence on S\'{e}rsic index $n$ at $M_{\ast}<10^{10.2}M_{\sun}$, while above this mass galaxies with a higher $n$ tend to be redder in the central NUV$-r$ color. These results suggest that galaxies with $M_{\ast}<10^{10.2}M_{\sun}$ exhibit similar star formation activity from the inner $R<R_{50}$ region to the $R>R_{50}$ region. In contrast, a considerable fraction of the $M_{\ast}>10^{10.2}M_{\sun}$ galaxies, especially those with a high $n$, have harbored a relatively inactive bulge component.

\end{abstract}
\keywords{galaxies: evolution -- galaxies: star formation}

\section{Introduction}
The issue of how star formation gets shutdown in galaxies (quenching) remains less understood to date. From the theoretically perspective, several quenching mechanisms have been proposed, including the active galactic nucleus (AGN) feedback \citep{Croton 2006, Hopkins 2006}, halo shock-heating \citep{Dekel 2006,Cattaneo 2006}, morphological quenching \citep{Martig 2009} and the environmental effects \citep{Gunn 1972,Toomre 1972,Moore 1996, Boselli 2006,Weinmann 2009}. However, observationally it is extremely difficult to identify the dominant working mechanism. For instance, massive galaxies preferentially reside in dense regions \citep{Li 2006}, .i.e., the halo quenching and environmental effects are possibly both at work. In the meanwhile, they have a high probability of simultaneously hosting an AGN and a bulge component \citep{Kauffmann 2003a,Heckman 2004, Schawinski 2010}, making it difficult to isolate the effect of AGN feedback from that of a central bulge. Such complexities have greatly hampered our knowledge of the detailed quenching picture.

In order to identify the dominant mechanism by which star formation shuts down in galaxies, much works explore the dependence of quiescence, or the quenched fraction upon a specific variable, while keeping other variables fixed \citep{Peng 2010,Bell 2012,Cheung 2012,Quadri 2012,Woo 2013,Kovac 2014,Bluck 2014, Lang 2014,Omand 2014,Knobel 2015, Woo 2015}. For instance, \citet{Cheung 2012} find that the $U-B$ color of a galaxy is mostly tightly linked to a dense galaxy inner structure. Similar results are reported by \citet{Lang 2014} and \citet{Bluck 2014}, who further show that quenched fraction is most tightly linked to bulge mass. Given the tight correlation between bulge mass and black hole mass, the authors suggest that AGN feedback is most favored.

Note that the above methodology strongly relies on the tightness of the measured quiescence--variable relation. This can be misleading in some cases. As shown in observations and simulations, a fraction of the quenched galaxies will undergo dry merging \citep{van Dokkum 2005, Naab 2007,Bundy 2009, Kormendy 2009,Tal 2009,Hopkins 2010}, through which the properties of galaxies are reshaped and the clues of quenching are smeared out. As a consequence, some properties of the dry merging remnants may be well correlated with quiescence, however they are not directly pointing to the quenching processes. An example is the phenomenology of "mass quenching" \citep{Peng 2010}, that galaxies with high stellar mass ($M_{\ast}$) tend to have a high quenched fraction. There have been evidences showing that the properties of giant quenched galaxies (such as structure and Faber--Jackson relation) are more reasonably interpreted in a dry merging framework \citep{Kormendy 2009, Bernardi 2011a,Bernardi 2011b,Kormendy 2013}.

Dry merging is obviously irrelevant to quenching. To avoid such a confusion, one could alternatively search for the clues at the epoch when a galaxy is not fully quenched yet. Such attempts have been carried out by studying the green valley galaxies \citep{Mendez 2011, Pan 2013,Schawinski 2014, Pan 2014}. To make a step further, it will be exiting to probe the onset of quenching in a star-forming galaxy (SFG). However, this is very challenging since it is difficult to determine which SFG is going to quench. A practicable way would be studying the star formation properties of individual SFGs to search for the early quenching signals. Ideally, a direct investigation of the specific star formation rate (sSFR) distribution of a SFG will be helpful in diagnosing whether quenching has occurred in that galaxy. Such studies require a data set which contains 2-dimensional information of individual galaxies, for example, the integral field spectroscopic (IFS) data. The IFS sample obtained in the early projects (such as SAURON \citep{Bacon 2001} and DiskMass \citep{Bershady 2010}) is proved to be powerful in studying galactic physics. However, the existed IFS sample is still relatively small. To collect a larger IFS sample, several large projects such as MaNGA \citep{Bundy 2015}, CALIFA \citep{Sanchez 2012} and SAMI \citep{Croom 2012} are launched recently. Exploiting the early data products of these projects, some studies have begun to study the 2-dimensional star formation histories of a large local galaxy sample \citep{Perez 2013, Cid 2013,Gonzalez 2014, Li 2015, Gonzalez 2015}.

To resolve the sSFR of a large galaxy sample, one can alternatively use the existed multi-band imaging data. The 2-dimensional properties of galaxies can also be studied by utilizing spatially-resolved multi-band photometry \citep{de Jone 1996, Kong 2000, Mu 2007,Suh 2010, Lin 2013}. In this paper, we use the $GALEX$ \citep{Martin 2005} and SDSS \citep{York 2000} data to resolve the star formation properties inside/outside the half light radius of a large local SFG sample.  Throughout this paper, we assume a concordance $\Lambda$CDM cosmology with $\Omega_{\rm m}=0.3$, $\Omega_{\rm \Lambda}=0.7$, $H_{\rm 0}=70$ $\rm km~s^{-1}$ Mpc$^{-1}$, and a \citet{Kroupa 2001} IMF.

\section{Method and data used}
This work makes use of the UV--optical multi-band data, which are drawn from the SDSS and $GALEX$ survey. With these data, it is still challenging to directly measure of the sSFR distribution of galaxies. To be simplified, in this work we will use NUV$-r$ color as the representative of sSFR. Since UV luminosity $L_{\rm UV}$ is an SFR indicator \cite[e.g.,][]{Kennicutt 1998},  NUV$-r$ is thus a proxy of SFR/$L_{\rm r}$, where $L_{\rm r}$ is the $r$-band luminosity. In this sense, NUV$-r$ is a luminosity weighted sSFR. \citet{Salim 2005} first establish that the star formation history of a galaxy can be well constrained on the basis of its NUV$-r$ color. In a recent paper, \citet{Salim 2014} further shows that NUV$-r$ is more tightly correlated with sSFR than $u-r$ and $g-r$, especially in the low sSFR regime.

The $\it GALEX$ imaging has a relatively low resolution (The \emph{GALEX} NUV image has a resolution of 1 pixel=1\farcs5 and a point spread function (PSF) with full width at half-maximum (FWHM)=5\farcs3), which hampers a high-resolution study for a large SFG sample. To investigate a large sample, in this work we only resolve the NUV$-r$ color of galaxies in two regions, .i.e, a central region and an outer region. We refer the central region to be $R_{\rm central}\approx R_{50}$, where $R_{50}$ is the radius enclosing 50\% of the SDSS $r$-band petrosian flux. By doing this, one can simply perform aperture photometry to obtain NUV$-r_{\rm central}$ for a galaxy sample which with a similar $R_{50}$ angular size.

Based on the UV--optical raw imaging data, we have generated a multi-band photometric catalog following the pipeline of \citet{Wang 2010}. Here we give a brief review of the data processing. As the first step, we cross-matched the SDSS Data Release 8 spectroscopic sample \citep{Aih11} with the $GALEX$ DR6 frames and downloaded their UV+opical images. For the FUV and UV bands, only those with an exposure time greater than 1000 seconds are finally used. We register every image to the frame geometry of the NUV band. The images of each galaxies are then sheared into $3\farcm 0\times3\farcm 0$ stamps. Since the SDSS image has a higher resolution than the UV image, we degrade the former by convolving a NUV PSF kernel so that the UV and optical images are finally spatially matched. The PSF kernel is generated by stacking the star images in that $GALEX$ frame. For the PSF-matched galaxies, we then measure their magnitudes over 5 apertures, with $r=$[1\farcs5, 3\farcs0, 6\farcs0, 9\farcs0, 12\farcs0] by running the software \texttt{SExtractor} \citep{Bertin 1996}. \texttt{SExtractor} also measures the total magnitude of a galaxy ($mag_{\rm auto}$). Finally, we correct the magnitudes for galactic extinction using the galactic dust map of \citet{Schlegel 1998}.

Some publicly available data are also utilized. The structure parameter catalog used in the following is from \citet{Simard 2011}. \citet{Simard 2011} perform a bulge+disk structure decomposition to the SDSS spectroscopic galaxies with the software \texttt{GIM2D}. The global galaxy S\'{e}rsic index $n$ used in this work is from the $r$-band modeling, through which $n$ is allowed to vary from 0.5 to 8.  Both stellar mass ($M_{\ast}$) and the 4000 \AA ~ break strength ($D_{n}4000$) are from the MPA/JHU database\footnote{http://www.mpa-garching.mpg.de/SDSS/DR7}. $M_{\ast}$ is derived following \citet{Kauffmann 2003b}, with a typical uncertainty of $\Delta~{\rm log}~M_{\ast}$=0.07 dex.
\begin{figure*}
\centering
\includegraphics[width=160mm,angle=0]{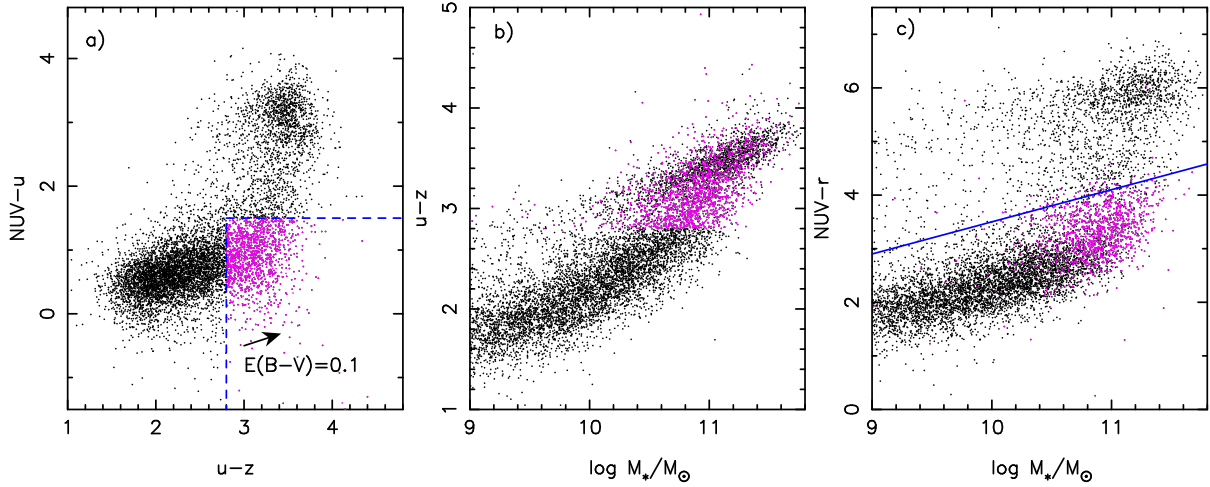}
\caption{a): the NUV$-u$ versus $u-z$ color--color diagram. It is clear that galaxies can be well separated into star-forming and quiescent ones in the two color diagram. However, in the star-forming sequence, a considerable fraction of the galaxies (denoted in pink symbols) have a relatively red $u-z$ color similar to the quiescent ones. The blue-dashed line encloses a region of $u-z>2.8$ and NUV$-u<1.5$. The arrow indicates the shift due to dust for an E(B--V)=0.1, assuming a dust attenuation law of \citet{Cardelli 1989}. b):  The $u-z$ versus $M_{\ast}$ diagram. It is clear that many dusty SFGs will be misclassified as quiescent galaxies when using this diagram. c) The NUV$-r$ versus $M_{\ast}$ diagram. Our final SFG selection criterion is shown in the blue line.}\label{fig1}
\end{figure*}

\begin{figure*}
\centering
\includegraphics[width=140mm,angle=0]{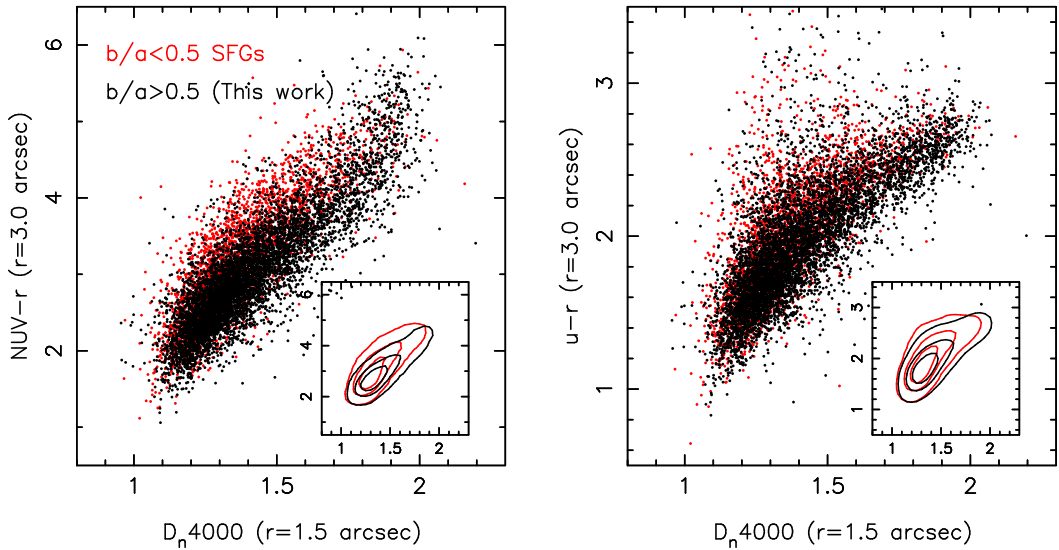}
\caption{Left: the correlation between NUV$-r_{r=3\farcs0}$ and the $D_{n}4000$ indices measured from the SDSS $r=1\farcs5$ fiber spectra. Face-on SFGs are shown in black symbols, while the edge-on ones are shown in red symbols. The low-right panel shows the contour plot of the scattered data points. Right: the correlation between $u-r_{r=3\farcs0}$ and $D_{n}4000$. It is clear that NUV$-r$ is a superior photometric indicator of $D_{n}4000$ than $u-r$.}\label{fig2}

\end{figure*}
\section{Sample selection}

Our UV-optical matched catalog contains 222,065 galaxies. To ensure that the measured central NUV$-r$ color is not significantly affected by the relatively poor resolution of the \emph{GALEX} images, we limit the sample galaxies to have $R_{50}\sim 6\farcs0 $, which is approximately 2 times the PSF size. With this size limit, we can simply take the NUV$-r$ measured inside the $R$=6\farcs0 aperture as a proxy of NUV$-r_{\rm R<R_{50}}$. In the following, NUV$-r_{\rm central}$ refers to NUV$-r_{R=6\farcs0}$ if not specifically stated. The detailed sample selection criteria are as follows:

 (1) $z=[0.01, 0.1]$, where $z$ is the SDSS spectroscopic redshift. 88,465 out of the 222,065 galaxies are within this redshift range.

 (2) minor$-$major axis ratio $b/a>0.5$. NUV$-r$ color is very sensitive to dust attenuation. Removing the edge on objects can greatly alleviate the dust reddening effect on the color measurement. Besides dust effect, the central region of an edge-on galaxy is a superposition of its disk and bulge component, making it difficult to derive its true central color. Thus we remove the edge on galaxies from further analysis. This criterion keeps 59,177 out of the 88,465 ones that pass (1).

 (3) 4\farcs0<$R_{50}$<8\farcs0, to ensure that the $r=6\farcs0$ aperture encloses $\sim 50$\% of the total flux of a galaxy.
This criterion is quite strict and only 9,391 out of the 59,177 galaxies are passed.

 (4) We restrict the galaxies to be brighter than 23.0 mag in the NUV band, to ensure the accuracy of the color measurement. In fact, as shown bellow, our selected SFGs mostly have $m_{\rm NUV}$<20.0 mag. Of the 9,391 galaxies, 8,980 were kept.

 (5) stellar mass $M_{\ast}>10^{9.0}~M_\sun$.

Criteria of (1)$-$(5) finally yield a sample of 8,200 galaxies. We note that this sample is not volume-completed and our conclusions do not rely on the selection completeness in the volume.

The observed colors of galaxies are more or less reddened by their dust contents. To separate the dust-reddened SFGs from truely passive galaxies, a color--color diagram is commonly adopted in the literature \citep{Williams 2009, Bell 2012,Chang 2015}. In panel a) of Figure~\ref{fig1}, we plot the selected 8,200 galaxies on the $NUV-u$ versus $u-z$ color--color diagram. One can find that a significant fraction of the galaxies have $u-z$ colors that are as red as those of the passive galaxies, whereas lie on the star-forming sequence, i.e., they are still SFGs. In panel b), it is clear many optically red SFGs will be misclassified as passive galaxies when using the optical color--mass diagram. Note that most of these galaxies have log($M_{\ast}/M_{\sun}$)>10.4. Panel c) shows the NUV$-r$ versus $M_{\ast}$ diagram. On this diagram, one can see that the optically red SFGs still lie on the star forming sequence. Our final SFG selection criterion is defined as NUV$-r$=0.6$\times$log$M_{\ast}$--2.5, which is shown in the blue line of panel c). This criterion is approximately equivalent to the color--color selection and will not miss the dusty SFGs. The final SFG sample contains 6,324 galaxies.

\begin{figure*}
\centering
\includegraphics[width=160mm,angle=0]{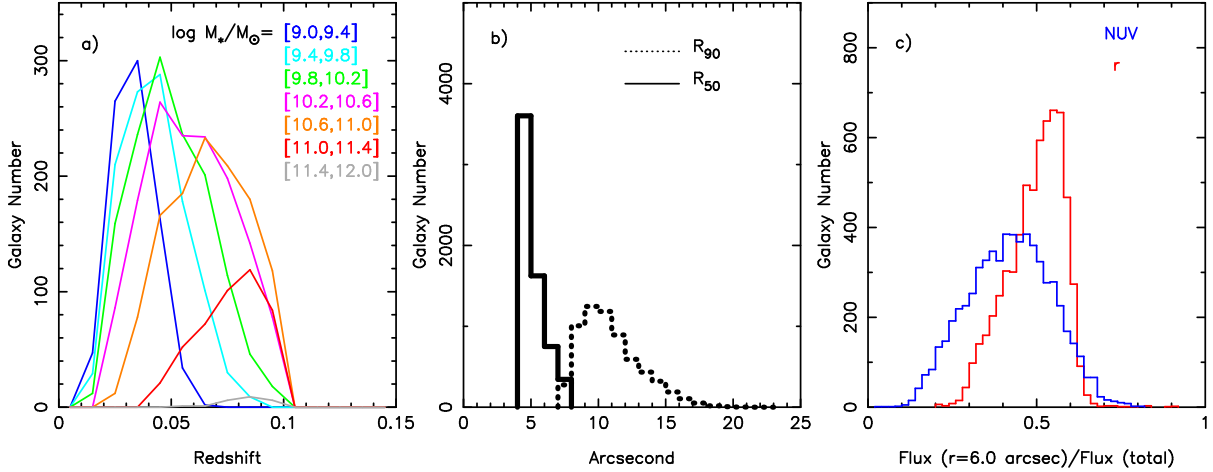}
\caption{a): the redshift distribution of the SFG sample. The SFG sample is divided into 7 subsamples according to stellar mass and each subsample is coded with a specific color. b): the $R_{50}$ and $R_{90}$ distributions of SFGs, shown in the solid line and the doted line, respectively. Both $R_{50}$ and $R_{90}$ and shown in angular size but not in physical size. c):the $f_{\rm central}$ distribution of SFGs, where $f_{\rm central}$ was defined as $f_{\rm central}=\rm Flux_{\rm r=6\farcs0}/\rm Flux_{\rm total}$. The $\emph{GALEX}$ NUV and SDSS $r$-band is shown in blue and red histograms, respectively..  }\label{fig3}
\end{figure*}

Although the good correlation between NUV$-r$ and sSFR has been established in \citet{Salim 2014}, at this point it is still necessary to investigate to what extend NUV$-r$ can trace sSFR for our sample. We explore the relation between NUV$-r$ and the 4000 \AA ~break strength ($D_{n}$4000). $D_{n}$4000 is in good correlation with sSFR and usually taken as a mean stellar age indicator \citep{Brinchmann 2004}. More importantly, $D_{n}$4000 is dust-free. In Figure~\ref{fig2}, we compare the NUV$-r$ measured in a central $r=3\farcs 0$ aperture with $D_{n}$4000 \footnote{$D_{n}$4000 is measured from the SDSS $r=1\farcs5$ fiber spectra.}. We use NUV$-r_{r=3\farcs 0}$ due to the resolution of $GALEX$ imaging. For comparison, we also plot the SFGs that with $b/a<0.5$ on this diagram. As shown in the figure, NUV$-r$ and $D_{n}$4000 form a linear correlation. After excluding the edge on galaxies, the tightness of this correlation is significantly improved. In the right panel, we show the correlation between $u-r$ and $D_{n}$4000. It is clear that $u-r$ is less sensitive to $D_{n}$4000 when $D_{n}$4000>1.6, because optical colors are insensitive to low level sSFR.

Note that in Figure~\ref{fig2}, NUV$-r$ and $D_{n}$4000 are measured over different apertures. If $D_{n}$4000 exhibits a non-zero gradient within the $r=3\farcs 0$ aperture, then it will bring some scatters into this relation. \citet{Li 2015} show that centrally quenched galaxies usually exhibit significantly negative $D_{n}4000$ gradient over $R\sim R_{e}$, where $R_{e}$ is the effective radius. In contrast, centrally star-forming ones do not have obvious gradient. One can find that the correlation seems shift rightward at $D_{n}$4000>1.6 than the outward extrapolation of the $D_{n}$4000<1.5 data points, which is expected according to the findings of \citet{Li 2015}. Once NUV$-r$ and $D_{n}$4000 are taken from the same region, the correlation between them should be even better than that shown in panel a). In general, Figure~\ref{fig2} supports that for the face-on objects, NUV$-r$ is a good photometric sSFR (or stellar age) indicator.

In panel a) of Figure~\ref{fig3}, we show the redshift distribution of the 6,324 SFG sample. As can be seen, for galaxies of fixed $M_{\ast}$, only those within a certain redshift interval are finally selected. This is due to the combination of the underlying $M_{\ast}$--size relation and criterion (3). Panel b) shows the $R_{50}$ and $R_{90}$ distributions for our sample. Both $R_{50}$ and $R_{90}$ are shown in angular size but not in physical size. We expect the SFGs to have a $R_{50}$ distribution which peaks at $R_{50}\sim$ 6\farcs0, so that the $r=6\farcs0$ aperture can well corresponds to the half light radius of this sample. From panel b) one finds that most of the SFGs have an angular size of $R_{50}\sim 5\farcs 0$. Panel d) shows the central flux fraction ($f_{\rm central}=flux_{\rm central}/flux_{\rm total}$) distribution. Here the total flux is converted from $mag_{\rm auto}$. One can see that the central flux fraction peaks at $f_{\rm central}\sim 0.55$ in the $r-$band, which is slightly higher than our expectation. This is due to the relatively low $R_{50}$ distribution of this sample. However, the $f_{\rm central}$ distribution is quite narrow, which mainly spans from 0.4 to 0.6. Thus in the following analysis one can still treat $R_{\rm central}$ roughly as $R_{50}$.

\section{The uncertainty of NUV$-r$}
\begin{figure}
\centering
\includegraphics[width=80mm,angle=0]{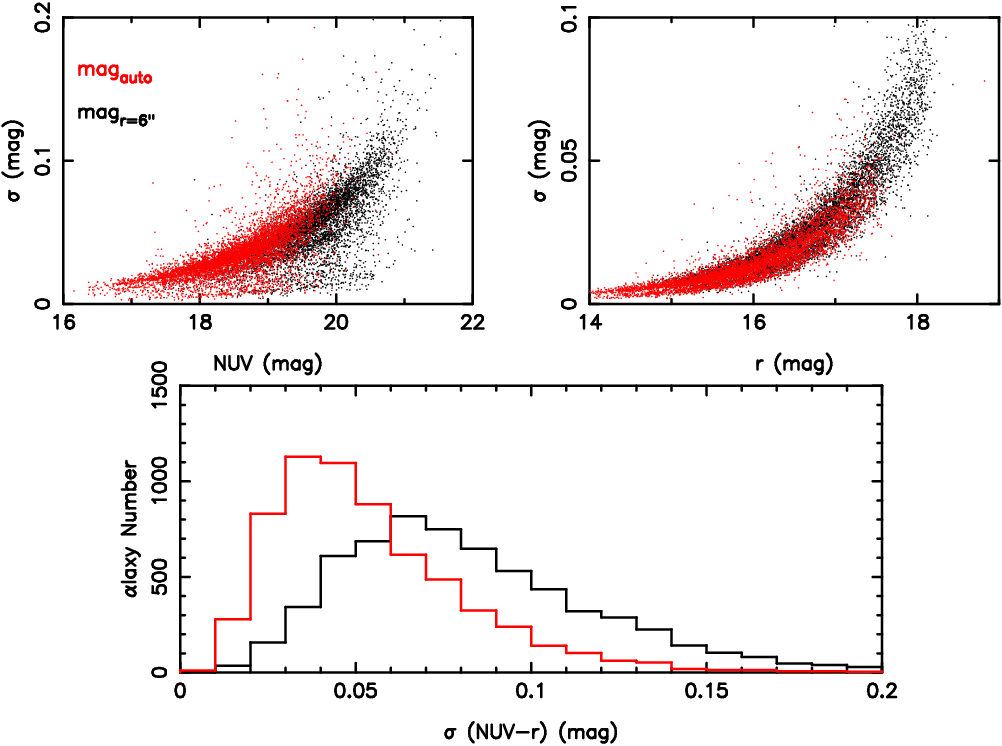}
\caption{Top left: the magnitude uncertainty as a function of magnitude in the $GALEX$ NUV band. Black symbols denote those measured in the central $r=6\farcs0$ aperture, while red symbols denote those of the total galaxy. Top right: similar to top left, but in the SDSS $r-$band. Bottom: the NUV$-r$ uncertainty distribution. }\label{fig4}
\end{figure}

In the next section we will carry out an analysis on NUV$-r_{\rm central}$ and NUV$-r_{\rm outer}$. Before doing this it is important to assess the accuracy of our color measurement. Figure~\ref{fig4} presents the output magnitude uncertainty measured by \texttt{SExtractor} for our SFGs.  Measurements of the $r=6\farcs0$ aperture and the total galaxies are shown in black and red symbols, respectively. In general, fainter objects have greater magnitude uncertainties. One can see that most SFGs have $mag_{\rm auto}$ <20.0 mag in the NUV band, which is $\sim$3.0 mag brighter than the NUV limiting magnitude. The color uncertainty distributions are shown in the bottom panel. In this panel one can see that the uncertainties of the NUV$-r_{\rm central}$ and NUV$-r_{\rm total}$ are $\sim$ 0.08 mag and 0.04 mag, respectively. Assuming that the color uncertainty of NUV$-r_{\rm outer}$ is of a similar order to that of NUV$-r_{\rm central}$, the uncertainty of the color discrepancy between the two probed sub-regions, $\sigma$[(NUV$-r_{\rm central}$)--(NUV$-r_{\rm outer}$)], is $\sim$ 0.15 mag.

\section{result}

\subsection{The dependence of NUV$-r_{\rm central}$ and NUV$-r_{\rm outer}$ on $M_{\ast}$}

\begin{figure*}
\centering
\includegraphics[width=160mm,angle=0]{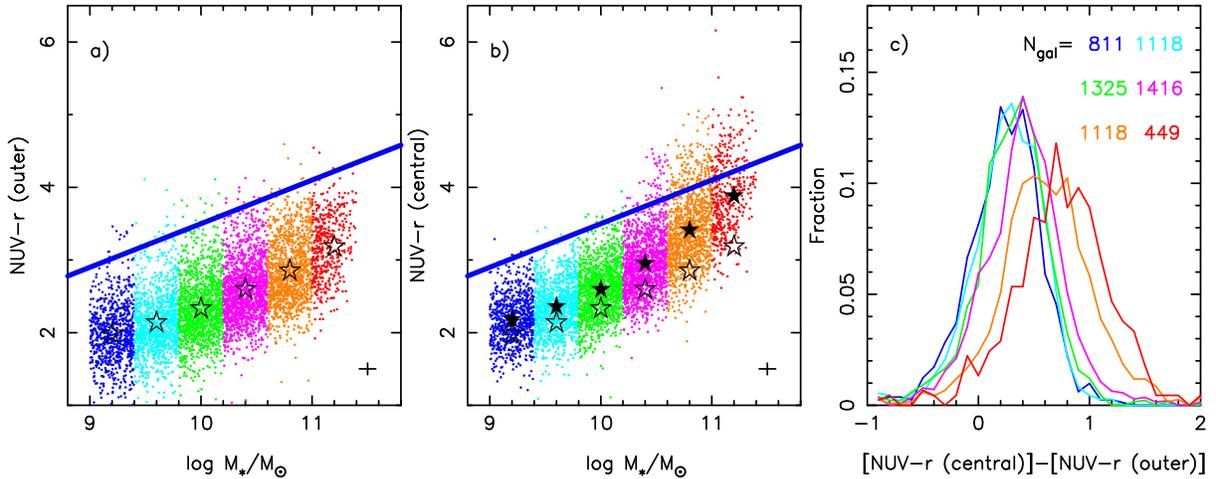}
\caption{a): NUV$-r_{\rm outer}$ as a function of $M_{\ast}$. Galaxies are binned in 6 mass bins, as shown in different colors. The open large symbols denote the median NUV$-r_{\rm outer}$ in that mass bin. The blue line shows our SFG selection criterion. b): similar to a), but shown in NUV$-r_{\rm central}$. The median value of NUV$-r_{\rm central}$ is shown in large solid symbols. For comparison, we also plot the median value of NUV$-r_{\rm outer}$.  c) the (NUV$-r_{\rm central}$)--(NUV$-r_{\rm outer}$) distributions of the 6 mass bins. In each mass bin, the galaxy number is marked. }\label{fig5}
\end{figure*}

Panel a) of Figure~\ref{fig5} explores the relation between NUV$-r_{\rm outer}$ and $M_{\ast}$. The SFGs are divided into 6 mass bins, with a bin size of $\Delta~{\rm log}M_{\ast}$=0.4 dex. As one can see,  NUV$-r_{\rm outer}$ becomes redder when increasing $M_{\ast}$.  Panel b) shows the (NUV$-r_{\rm central}$)--$M_{\ast}$ relation. For comparison, we overplot the (NUV$-r_{\rm outer}$)--$M_{\ast}$ relation as well. It is clear that NUV$-r_{\rm central}$ is only slightly redder than NUV$-r_{\rm outer}$ at log($M_{\ast}/M_{\sun}$)<10.2. Above log($M_{\ast}/M_{\sun}$)=10.2, the discrepancy between the two colors becomes larger towards higher $M_{\ast}$.

Panel c) shows the (NUV$-r_{\rm central}$)--(NUV$-r_{\rm outer}$) distributions. Note that we do not apply any internal extinction correction to both NUV$-r_{\rm central}$ and NUV$-r_{\rm outer}$. If there were no dust, (NUV$-r_{\rm central}$)--(NUV$-r_{\rm outer}$) is a direct tracer of the sSFR discrepancy of the two regions. However, SFGs contain dust and dust attenuation has been shown to be a strong function of $M_{\ast}$ \citep{Reddy 2010,Whitaker 2012, Whitaker 2014}. For this work, understanding the dust maps of galaxies will be helpful in interpreting the presented results. \citet{Gonzalez 2015} explore the 2-dimensional properties of $\sim$ 300 galaxies using the CALIFA IFS data. They find that the stacked $A_{V}$ profile shows a significant negative gradient in the innermost regions of galaxies. Beyond $R\sim0.5R_{50}$, the $A_{V}$ profile becomes flat. \citet{Nelson 2015b} use the 3D-HST grism data to investigate the dust maps of SFGs at $z\sim$1.4. They find that low mass SFGs have little dust attenuations at all radii. Above log($M_{\ast}/M_{\sun}$)$\sim$10.0, the dust attenuation has a significant negative gradient in the inner $R<$1.0 kpc region. Beyond $R\sim 1$ kpc, dust attenuation does not vary much. The findings of \citet{Gonzalez 2015} and \citet{Nelson 2015b} indicate that in the galaxy scale, the dust attenuation is not a strong function of radius. A similar conclusion is also reached by \citet{Ig 2013}. Given this, NUV$-r_{\rm central}$ might be only slightly dust-reddened than NUV$-r_{\rm outer}$.

An interesting feature presented in panel c) is that (NUV$-r_{\rm central}$)--(NUV$-r_{\rm outer}$) all peaks at $\sim$0.25 mag for the log($M_{\ast}/M_{\sun}$)<10.2 SFGs. After accounting for dust correction, the intrinsic color discrepancy should be even smaller. This indicates that the central regions of low mass SFGs are still actively forming new stars.  Above log($M_{\ast}/M_{\sun}$)=10.2, the peaks of the distributions shift redward and show a dependence on $M_{\ast}$. Since we have excluded the edge on SFGs to minimize dust effects on colors, a red central color strongly indicates an old bulge stellar population. Our result is broadly consistent with \citet{Nelson 2015a}, who find that the stacked sSFR profile of the $z\sim1$ SFGs is quite flat at log($M_{\ast}/M_{\sun}$)$<$10.0, whereas it exhibit a remarkable negative gradient for more massive galaxies (see their Figure 12). If the presence of old bulge components in massive SFGs is confirmed, it naturally explains the flattening of the SFR--$M_{\ast}$ relation of SFGs seen at the massive end towards low $z$ \citep{Whitaker 2014,Schreiber 2015,Lee 2015,Gavazzi 2015}. Our finding also provides a window to interpret the intrinsic scatter of the star formation main sequence (MS) of local SFGs. Recently, \citet{Guo 2013} and \citet{Guo 2015} find that the intrinsic scatter of the low-z MS increases with $M_{\ast}$, which is not reported in other similar studies \cite[e.g.,][]{Whitaker 2012}. In the context of this work, Guo et al's findings seem to be more reasonable since many massive SFGs have been partly quenched (see Figure~\ref{fig5}).

\begin{figure*}
\centering
\includegraphics[width=110mm,angle=0]{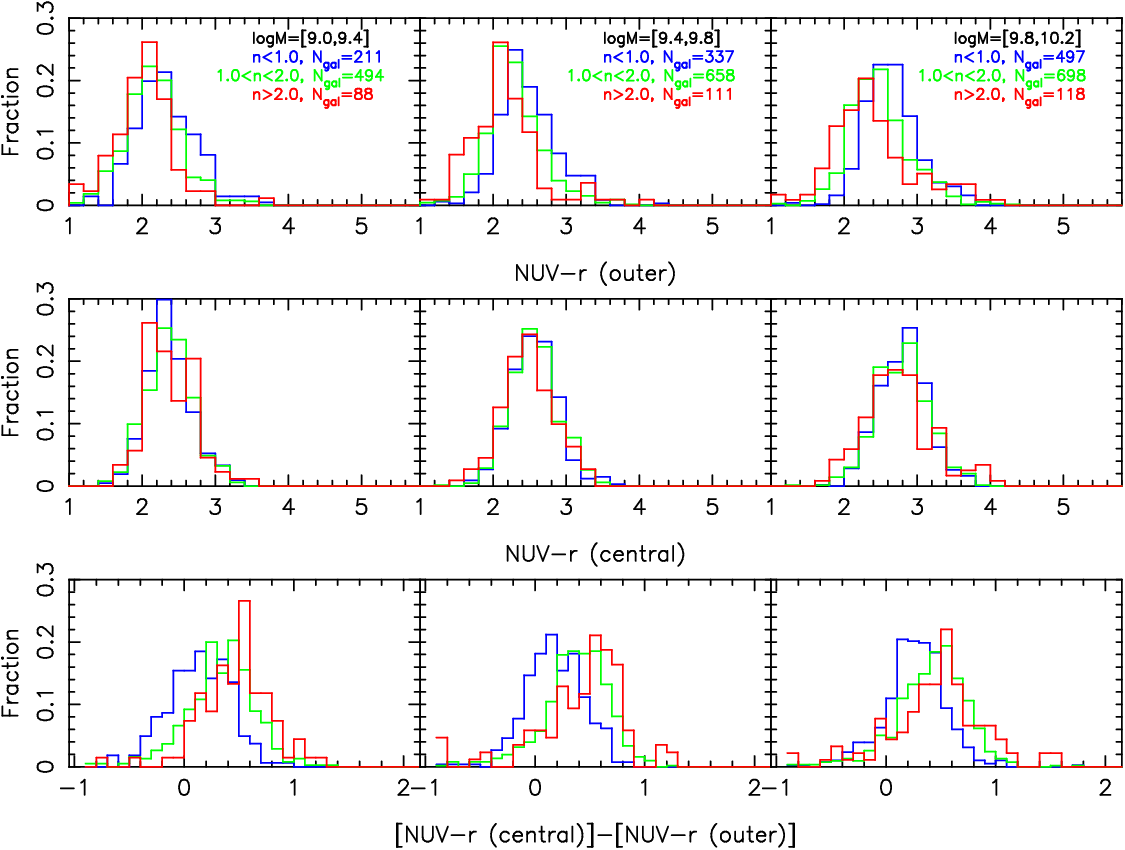}
\caption{Top panels: the NUV$-r_{\rm outer}$ distributions. From left to right, we show the results for the log($M_{\ast}/M_{\sun}$)=$[9.0,9.4]$, log($M_{\ast}/M_{\sun}$)=$[9.4,9.8]$ and log($M_{\ast}/M_{\sun}$)=$[9.8,10.2]$ subsamples, respectively. Galaxies within different $n$ bins are shown in different colors. The number of galaxies are also marked in each panel. Middle panels: the NUV$-r_{\rm central}$ distributions. Bottom panels: the color discrepancy distributions. }\label{fig6}

\end{figure*}

\begin{figure*}
\centering
\includegraphics[width=110mm,angle=0]{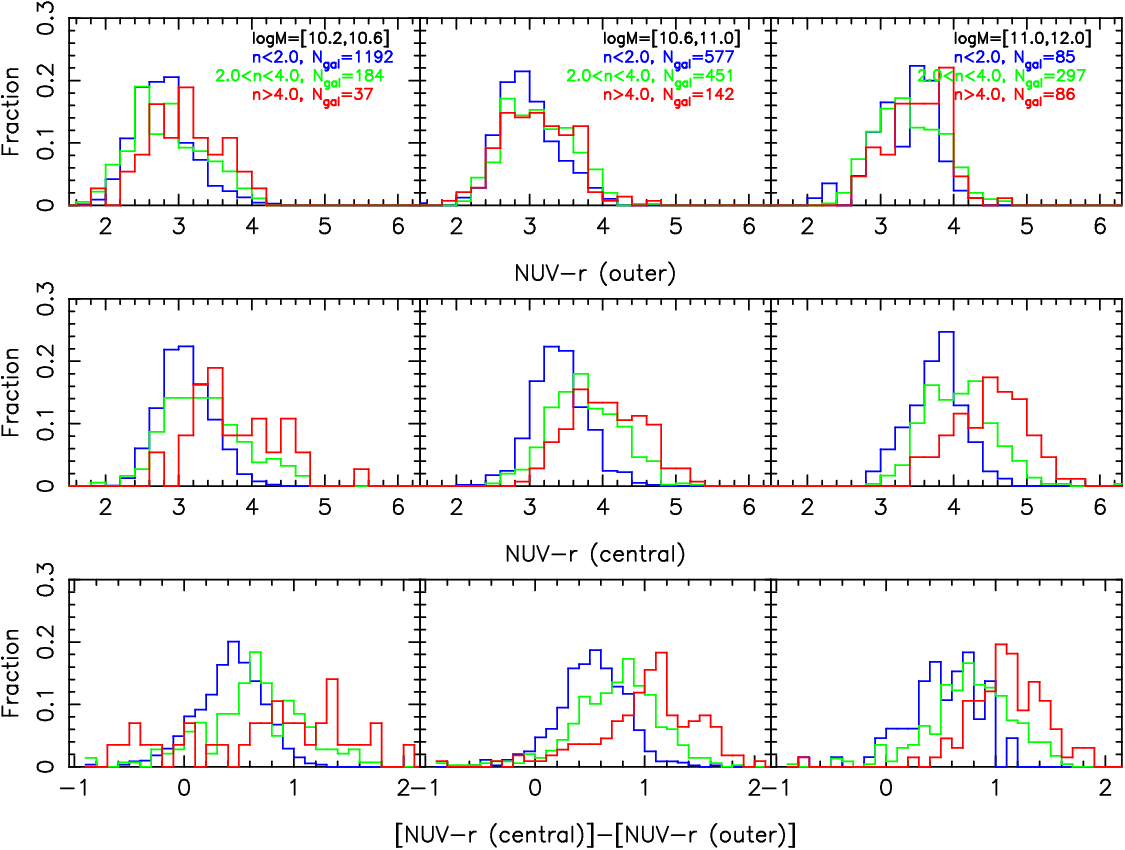}
\caption{Similar to Figure~\ref{fig6}, but for the SFGs with log($M_{\ast}/M_{\sun}$)>10.2 }\label{fig7}

\end{figure*}

\subsection{The dependence of NUV$-r_{\rm central}$ and NUV$-r_{\rm outer}$ upon galaxy structure}
Recent processes show that quiescence is well correlated with the presence of a dense galaxy inner structure \citep{Bell 2012, Cheung 2012, Fang2013, Woo 2015}. Given this, it will be interesting to explore whether bulgy SFGs have already shown some signs of early quenching. Galaxy structure can be parameterized by the bulge-to-total ratio ($B/T$), or S\'{e}rsic index $n$. The petrosian concentration index, parameterized by $C=R_{90}/R_{50}$, is also prevalently adopted in the early SDSS studies \citep{Strateva 2001,Kauffmann 2003b, Brinchmann 2004, Li 2006}. In the present work we do not attempt to investigate the dependence of our measured NUV$-r$ color on all the mentioned structural parameters. We have assessed which structure parameter is best tracing the early quenching of SFGs. Assuming that the quenching processes of massive galaxies start from their central regions, we select a massive SFGs sample (which with log($M_{\ast}/M_{\sun}$)>10.5) to investigate the dependence of their inner stellar age indicator (.i.e., the central $D_{n}4000$) upon $B/T$, $n$ and $C$. We found that the central $D_{n}4000$ is most strongly dependent on $n$ (Pan et al. in prep). Thus in this section we only explore the dependence of NUV$-r$ upon S\'{e}rsic index $n$.

In Figure~\ref{fig6} and Figure~\ref{fig7}, we show how NUV$-r_{\rm central}$ and NUV$-r_{\rm outer}$, as well as (NUV$-r_{\rm central}$)--(NUV$-r_{\rm outer}$) depend on $n$. Figure~\ref{fig6} shows the results for the log($M_{\ast}/M_{\sun}$)<10.2 SFGs. Galaxies are binned according to $n$. In the mass regime of log($M_{\ast}/M_{\sun}$)<10.2, one can see that NUV$-r_{\rm outer}$ is somewhat bluer for the high $n$ galaxies. In the middle panels, it is interesting that NUV$-r_{\rm central}$ seems largely independent of $n$. Figure~\ref{fig7} shows the results of the log($M_{\ast}/M_{\sun}$)>10.2 galaxies. As can be seen, NUV$-r_{\rm outer}$ is not strongly dependent on $n$. However, NUV$-r_{\rm central}$ shows a strong dependence on $n$, in the sense that the high $n$ galaxies have redder central colors.

\begin{figure*}
\centering
\includegraphics[width=140mm,angle=0]{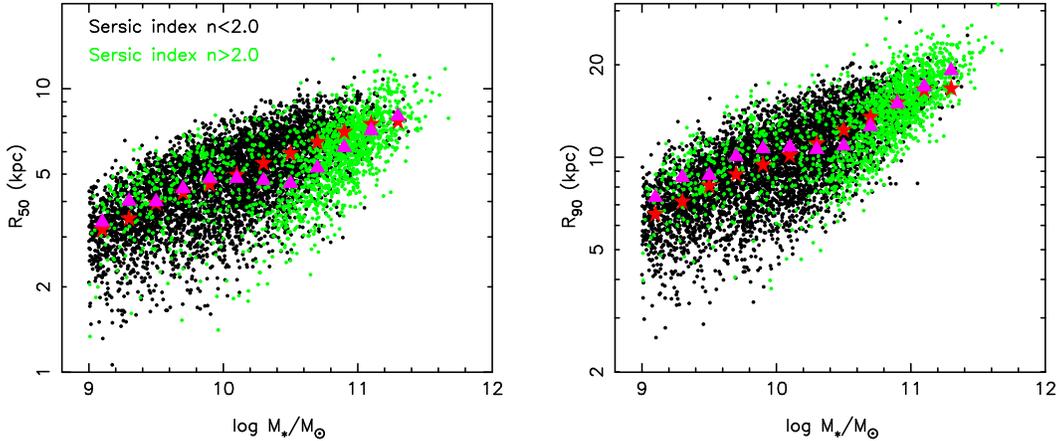}
\caption{Left: the $R_{50}$--$M_{\ast}$ relation. SFGs with $n$<2.0 and $n$>2.0 are shown in black and green small symbols, respectively. The median values of the $n$<2.0 and $n$>2.0 subsamples are shown in red and pink large symbols, respectively. Right: the $R_{90}$--$M_{\ast}$ relation. }\label{fig8}

\end{figure*}

Understanding the meaning of $n$ is important in interpreting the results presented in Figure~\ref{fig6} and Figure~\ref{fig7}. The S\'{e}rsic index $n$ describes the light/mass profile of a galaxy. Normally a high $n$ indicates a large $B/T$. However, in a bulge+disk system, the best-fit $n$ will increase either by placing more light to the bulge, or by growing the disk extent at a fixed bulge size (see \citet{Lang 2014} for a detailed discussion). In the former case, a high $n$ indicates a large bulge, while in the latter it indicates a very extended disk component. To distinguish between these two cases, we have investigated the $R_{50}$--$M_{\ast}$ and $R_{90}$--$M_{\ast}$ relations of the SFGs. The results are shown in Figure~\ref{fig8}. We find that at log($M_{\ast}/M_{\sun}$)<10.2, those galaxies with $n$>2.0 tend to have larger $R_{90}$ but similar $R_{50}$ compared to those with $n$<2.0, i.e., the low mass, high $n$ SFGs tend to have a very extended disk component (at least for our sample). For SFGs with log($M_{\ast}/M_{\sun}$)>10.2, we find that the $n$>2.0 galaxies tend to have smaller $R_{50}$ but similar $R_{90}$ than those with $n$<2.0. This supports that massive high $n$ galaxies tend to contain a prominent bulge component.

\citet{Wang 2011} find that a H~{\sc{i}} rich galaxy tend to have a larger and bluer outer disk than the mean. According to Figure~\ref{fig8} and \citet{Wang 2011}, it is not surprising to see that the low mass, high $n$ galaxies have relatively bluer NUV$-r_{\rm outer}$. In a recent work, \citet{Erf 2016} also find that the low mass, high $n$ SFGs have higher SFR than the mean (see their Figure 9), for which they have gave no explanation. At the massive end, they also found a dependence of SFR on $n$ similar to ours. \citet{Whitaker 2015} also suggest that the massive, high $n$ SFGs give rise to the flattened slope of the star formation main sequence at $z<1.0$. In this work, our spatially-resolved NUV$-r$ color enables a direct interpretation of the connections between star formation and $n$ in different mass regimes.

\section{Discussion}
\subsection{Is the red NUV$-r_{\rm central}$ due to heavily dust obscuration?}
The NUV$-r$ color is a good proxy of sSFR.  However, all color indices are more or less affected by the metallicity and dust of that galaxy. It is thus necessary to assess the impacts of metallicity and dust on our results. In Figure~\ref{fig5}, we draw that nearly all galaxies have their NUV$-r_{\rm central}$ and NUV$-r_{\rm outer}$ bellow 5.0. Over this range, NUV$-r$ is mainly determined by sSFR (or stellar age), rather than metallicity \citep{Kaviraj 2007a, Kaviraj 2007b}. Thus metallicity should has a minor effect on our results. However, dust attenuation is known to be more serious towards the massive end. Are the results of Figure~\ref{fig5} and Figure~\ref{fig7} caused by heavily dust attenuation of massive SFGs?

\begin{figure}
\centering
\includegraphics[width=80mm,angle=0]{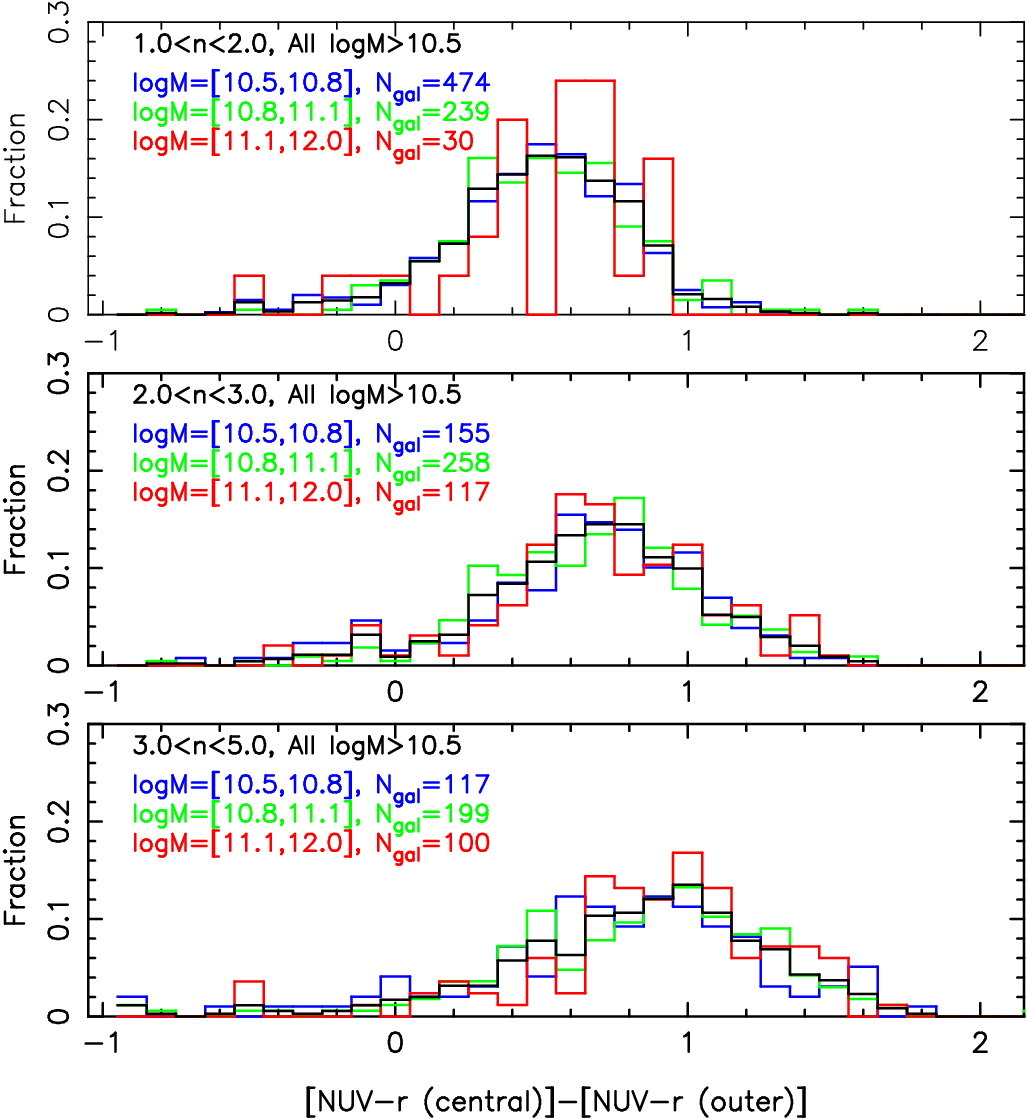}
\caption{The color discrepancy distributions for log($M_{\ast}/M_{\sun}$)>10.5 SFGs. From top to bottom, we divide the sample into 3 $n$ bins. Galaxies with different $M_{\ast}$ are shown in different colors.  }\label{fig9}

\end{figure}

To access the role of dust on our results, we explore the (NUV$-r_{\rm central}$)--(NUV$-r_{\rm outer}$) distributions as a function of $M_{\ast}$ within the range of log($M_{\ast}/M_{\sun}$)=$[10.5,12.0]$. The results are presented in three fixed $n$ bins, as shown in Figure~\ref{fig9}. It can be seen that when keeping $n$ fixed, the color discrepancy distribution shows a weak dependence on $M_{\ast}$. However, when comparing the results between different $n$ bins, we find that these distributions show a clear dependence on $n$, in the sense that high $n$ SFGs tend to have red (NUV$-r_{\rm central}$)--(NUV$-r_{\rm outer}$) distributions. The insensitivity of (NUV$-r_{\rm central}$)--(NUV$-r_{\rm outer}$) on $M_{\ast}$ is not expected if dust produces the results of Figure~\ref{fig5}.

An additional supporting evidence comes from the uniformity of the (NUV$-r_{\rm central}$)--(NUV$-r_{\rm outer}$) distributions bellow log($M_{\ast}/M_{\sun}$)=10.2. Previous works support that the dust attenuation of SFGs (traced by $L_{\rm IR}/L_{\rm UV}$) has significantly increased from log($M_{\ast}/M_{\sun}$)=9.0 to log($M_{\ast}/M_{\sun}$)=10.2 \citep{Whitaker 2012,Guo 2013, Whitaker 2014}. However, one draw from Figure~\ref{fig5} that the (NUV$-r_{\rm central}$)--(NUV$-r_{\rm outer}$) distributions all peak around $\sim$ 0.25 and with a similar scatter for the log($M_{\ast}/M_{\sun}$)<10.2 SFGs, indicating that the (NUV$-r_{\rm central}$)--(NUV$-r_{\rm outer}$) distribution is insensitive to the dust amount of SFGs. To conclude, we suggest that the red NUV$-r_{\rm central}$ colors of massive SFGs are owing to old bulge stellar populations, although the dust effect can not be fully ruled out. Future studies based on large IFS samples will shed more light on this topic.

\begin{figure}
\centering
\includegraphics[width=80mm,angle=0]{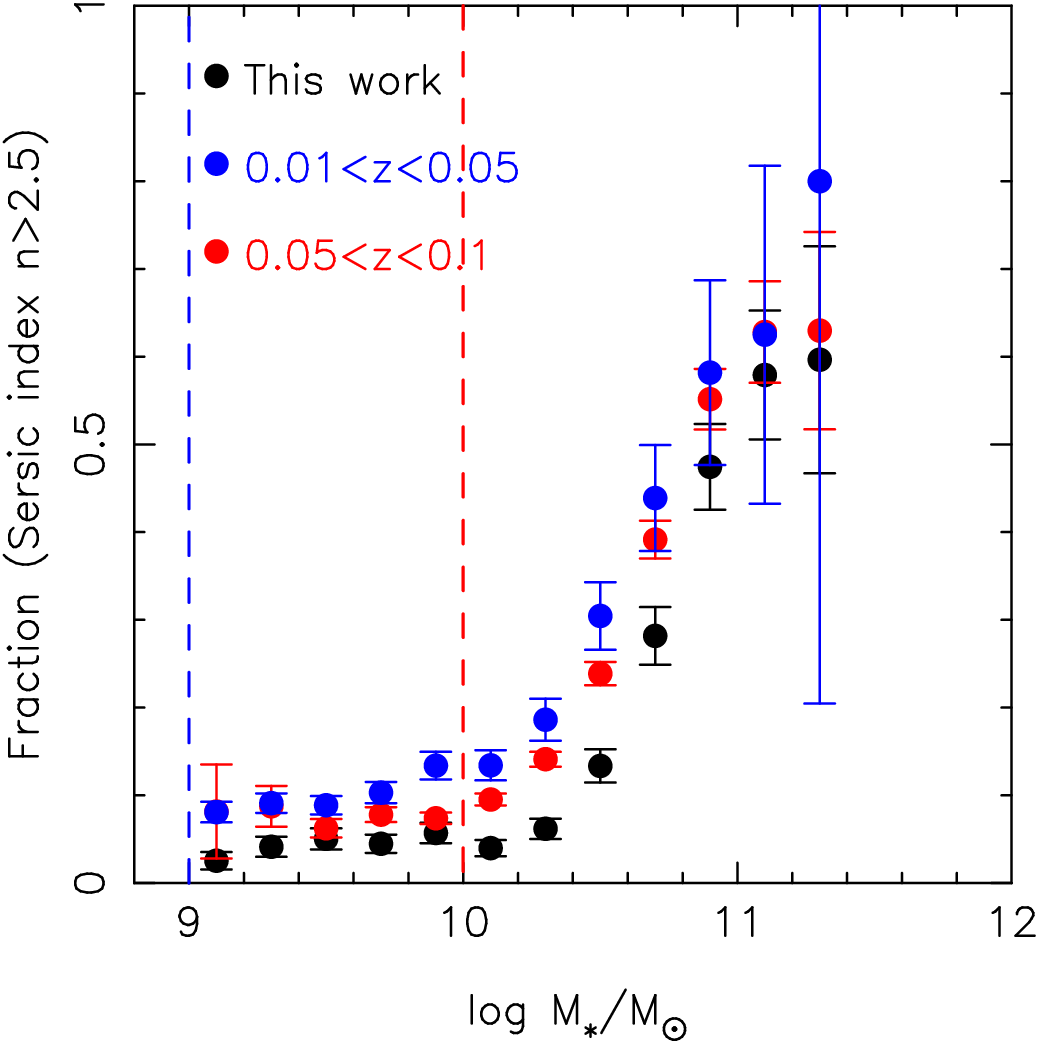}
\caption{The fraction of bulgy SFGs (with $n$>2.5) as a function of $M_{\ast}$. Data points are calculated in the $M_{\ast}$ bin size of $\Delta M_{\ast}$=0.2 dex. The red and blue are from the comparison samples within 2 different redshift bins. The red and blue dashed-vertical lines denote the mass limit of the comparison samples. }\label{fig10}

\end{figure}

\subsection{bulge growth and star formation quenching}
AGN feedback has long been served as a candidate mechanism of ceasing star formation in models \cite[e.g.,][]{Croton 2006,Hopkins 2006,Somerville 2008}. Observationally, there have been growing evidences supporting this scenario \citep{Schawinski 2006, Page 2012,Fabian 2012, Mai 2012,Barger 2015}. Recent progresses have confirmed the tight correlation between quiescence and the presence of a dense galaxy inner structure, whatever parameterized by $n$ \citep{Bell 2004,Driver 2006, Bell 2012, Cheung 2012}, by the inner 1 kpc stellar mass density \citep{Cheung 2012, Fang2013,Woo 2015} or by bulge mass \citep{Bluck 2014, Lang 2014}. Given the tight correlations between black hole mass, bulge mass and $n$, this observation supports the AGN feedback paradigm at a qualitative level. Figure~\ref{fig7} shows that massive, high $n$ SFGs indeed have relatively red NUV$-r_{\rm central}$ distributions, supporting that the star formation of bulge is at least partly controlled by the bulge-related process. However, this trend does not hold for the less massive galaxies (see Figure~\ref{fig6}). Figure~\ref{fig6} thus suggests that the  quenching picture of low mass SFGs is not the same as that of the massive ones. As already established, low mass galaxies are mainly quenched through environmental effects \citep{Peng 2010}. In a dense environment, gas stripping will cause quenching first on galactic outskirts, especially for the low mass systems \citep{Boselli 2014}. The "outside-in" quench mode of low mass galaxies has been recently confirmed by \citet{Pan 2015}.

\citet{Martig 2009} propose the build up of a central bulge is sufficient to stabilize the surrounding gas disk from collapsing and forming new stars, which can cause quenching in a SFG without expelling/depleting its gas. In a recent work, \citet{Guo 2015} find that the high $B/T$ SFGs have a larger intrinsic scatter in their disk sSFR, based on which they argued that the presence of a central bulge has a dramatically impact on the galaxy scale star formation. In this work, we show that the NUV$-r_{\rm central}$ of massive SFGs strongly depends on $n$, whereas NUV$-r_{\rm outer}$ does not (see Figure ~\ref{fig7}). This indicates that the disk star formation of a SFG is less affected by its inner structure. The "morphological quenching" scenario is thus not supported by our data.

Recently, \citet{Abramson 2014} suggest that a key part of "mass quenching" is the build up of an inactive bulge component. This is supported by Figure~\ref{fig7} of this work. Given that many massive, bulgy SFGs have been partly quenched, it is important to investigate the contribution of the "bulgy galaxies" to the whole SFG population. Figure~\ref{fig10} shows the fractions of "bulgy SFGs" (with $n$>2.5) as a function of $M_{\ast}$. For our sample, one can see that the this fraction is around 10\% at $M_{\ast}<10^{10.2}M_{\sun}$, while it rapidly increases to 60\%-- 70\% at $M_{\ast}\sim 10^{11.0}M_{\sun}$. A similar trend can be found in \citet{Fisher 2011} and \citet{Erf 2016}.

For comparison, we overplot the results of two SFGs samples drawn from $z=[0.01,0.05]$ and $z=[0.05,0.1]$. We use similar SFG selection criteria but do not apply any $R_{50}$ cut to the comparison samples. It can be seen that the results of the two comparison samples are well consistent with each other. One can also see that the fractions derived from the SFGs sample of this work are obviously lower than those of the comparison samples, especially at the mass range of log($M_{\ast}/M_{\sun}$)=[10.0,11.0]. We have checked the $R_{50}$ distributions of SFGs and confirmed that at fixed $M_{\ast}$ and $z$, SFGs in the highest $n$ quartile have a smaller $R_{50}$ distribution than the mean, especially for the massive ones, i.e., our $R_{50}$ cut is biased against high $n$ SFGs. Accounting for the sample selection bias, the intrinsic (NUV$-r_{\rm central}$)--(NUV$-r_{\rm outer}$) distributions should be even redder than those presented in Figure~\ref{fig5} at the massive end.

\section{Summary}
Aiming to resolve the sSFR of individual galaxies and to better understand how quenching process at its early phase, we measure the NUV$-r$ colors both inside and outside the half-light radius for a sample of 6,324 face-on local SFGs. A very tight linear correlation is found between NUV$-r$ and $D_{n}4000$, supporting that NUV$-r$ is a good photometric stellar age (or specific star formation rate) indicator, at least for these face-on SFGs. We investigate how these two colors depend on stellar mass and galaxy structure. Our findings are the following:

1. Bellow $M_{\ast} \sim 10^{10.2}M_{\sun}$, the central NUV$-r$ is on average only $\sim$ 0.25 mag redder than the outer NUV$-r$. The intrinsic value would be even smaller after accounting for dust correction; above $M_{\ast} \sim10^{10.2}M_{\sun}$, the central NUV$-r$ becomes systematically much redder than the outer NUV$-r$ for more massive galaxies.

2. Bellow $M_{\ast} \sim 10^{10.2}M_{\sun}$, the central NUV$-r$ shows no dependence on S\'{e}rsic index $n$. However, above this mass, galaxies with a higher $n$ tend to be redder in the central NUV$-r$ color. Our analysis suggests that the red central color presented in massive galaxies is owing to their old bulge stellar population.

Our findings thus suggest that the central regions of less massive SFGs are comparably active to the outer regions. In contrast, the bulge of a portion of massive galaxies, especially for those with a high $n$, has been quenched. The presence of old bulges in massive galaxies naturally explains the flattened slope of the SFR--$M_{\ast}$ relation seen at the massive end. Our findings also provide a direct interpretation of the intrinsic scatter of the star formation main sequence.

\acknowledgments
We thank the anonymous referee for insightful comments and a careful reading of the manuscript. This work was supported by the NSFC projects (Grant Nos. 11473053, 11121062, 11233005, U1331201, U1331110, 11225315, 1320101002, 11433005, and 11421303), the National Key Basic Research Program of China (Grant No. 2015CB857001), the ``Strategic Priority Research Program the Emergence of Cosmological Structures'' of the Chinese Academy of Sciences (Grant No. XDB09000000), the Specialized Research Fund for the Doctoral Program of Higher Education (SRFDP, No. 20123402110037), and the Chinese National 973 Fundamental Science Programs (973 program) (2013CB834900, 2015CB857004).

\end{document}